\documentclass[english,aps,floats,onecolumn,showpacs,nofootinbib]{revtex4}
\usepackage{pslatex}
\usepackage[T1]{fontenc}
\usepackage[latin1]{inputenc}
\usepackage{graphicx}
\usepackage{epsfig}

\usepackage{calc}
\usepackage{ifthen}

{
{
{
\newcommand{\bea}{\begin{eqnarray}}
\newcommand{\eea}{\end{eqnarray}}

\newcommand{\nc}{\newcommand}
\nc{\renc}{\renewcommand}
\nc{\eqs}[2]{\mbox{Eqs.~(\ref{#1},\,\ref{#2})}}
\nc{\eq}[1]{\mbox{Eq.~(\ref{#1})}}
\nc{\figs}[2]{\mbox{Figs.~(\ref{#1},\,\ref{#2})}}
\nc{\fig}[1]{\mbox{Fig~.(\ref{#1})}}
\nc{\be}[1]{\begin{equation} \mbox{$\label{#1}$}}
\nc{\ee}{\vspace{0.1cm}\end{equation}}

\newcommand{\bean}{\begin{eqnarray*}}
\newcommand{\eean}{\end{eqnarray*}}

\def\lae{\;^{<}_{\sim} \;}

\def\ha{\hat{a}}

\def\ta{\tilde{a}}

\begin{document}
\title{Signatures of Planck Corrections in a Spiralling Axion Inflation Model}
\author{John McDonald}
\email{j.mcdonald@lancaster.ac.uk}
\affiliation{Dept. of Physics, University of 
Lancaster, Lancaster LA1 4YB, UK}

\begin{abstract}

The minimal sub-Planckian axion inflation model accounts for a large scalar-to-tensor ratio via a spiralling trajectory in the field space of a complex field $\Phi$. Here we consider how the predictions of the model are modified by Planck scale-suppressed corrections. In the absence of Planck corrections the model is equivalent to a $\phi^{4/3}$ chaotic inflation model. Planck corrections become important when the dimensionless coupling $\xi$ of $|\Phi|^{2}$ to the topological charge density of the strongly-coupled gauge sector $F \tilde{F}$ satisfies $\xi \sim 1$. For values of $|\Phi|$ which allow the Planck corrections to be understood via an expansion in powers of $|\Phi|^{2}/M_{Pl}^{2}$, we show that their effect is to produce a significant modification of the tensor-to-scalar ratio from its $\phi^{4/3}$ chaotic inflation value without strongly modifying the spectral index. In addition, to leading order in $|\Phi|^2/M_{Pl}^{2}$, the Planck modifications of  $n_{s}$ and $r$ satisfy a consistency relation, $\Delta n_{s} = - \Delta r/16$. Observation of these modifications and their correlation would allow the model to be distinguished from a simple $\phi^{4/3}$ chaotic inflation model and would also provide a signature for the influence of leading-order Planck corrections.

\end{abstract}
 \pacs{}
 \maketitle

\section{Introduction} 

 There has recently been renewed interest in the theoretical issues surrounding a large tensor-to-scalar ratio $r$, following the claim by BICEP2 \cite{bicep1,bicep2} to have observed a large value for $r > 0.1$. Although it has since become clear that the dust signal in the analysis of \cite{bicep1,bicep2} was underestimated, the theoretical issues raised by the result remain interesting. A serious difficulty with large $r$ is the need for a super-Planckian value of the inflaton field in single-field inflation models. The Lyth bound \cite{lythb,antusch}
shows that the change of the inflaton field during inflation must be $\sim 10 M_{Pl}$ if $r \sim 0.1$\footnote{$M_{Pl} = 1/\sqrt{8 \pi G}$.}.  In this case additional non-renormalizable potential terms scaled by the Planck mass, which would be expected dimensionally, must somehow be suppressed. 

But even if the additive Planck potential corrections can be suppressed, in order to discuss inflation in the absence of a complete knowledge of Planck-scale physics it may still be necessary to have sub-Planckian field values during inflation. Moreover, it is possible that Planck corrections will present an insurmountable  barrier to inflation at super-Planckian field values. This is true in generic supergravity models \cite{nilles}, for example, where even if Planck-scale suppressed non-renormalizable superpotential terms can be eliminated (by an R-symmetry, for example), corrections due to the $e^K$ prefactor which multiplies the SUSY inflaton potential will still require that $|\Phi| < M_{Pl}$ in order that the potential can be calculated without knowledge of the complete 
K\"ahler potential $K$, or even to allow inflation to be possible should this prefactor grow exponentially once $|\Phi| > M_{Pl}$.

   Based on this example, we propose that, in the absence of a complete theory of Planck-scale physics, Planck corrections to the uncorrected inflaton potential $V_{0}(\phi)$ should generally be considered to consist of an  additive correction and a  multiplicative correction
\be{ex4} V = f\left( \frac{\phi}{M_{Pl}} \right) V_{0}(\phi) + g\left(\frac{\phi}{M_{Pl}} \right)   ~.\ee
In order to have a model of inflation which is robust with respect to Planck corrections, we should therefore (i) sufficiently suppress the additive term $g$ relative to the first term and (ii) ensure that the multiplicative term $f$ does not prevent inflation. As in the SUGRA example, where $g$ is due to Planck corrections to the superpotential whereas $f$ is due to the K\"ahler potential, $g$ can be independent of $f$. In the case where $\phi$ is close to $M_{Pl}$, we therefore need to assume that $g$ is no larger than $V_{0}$. But even with this assumption, in the absence of a complete Planck-scale theory giving the exact function $f$, we also need to assume that $\phi$ is sub-Planckian in order to be able to control $f$ via an expansion in $\phi/M_{Pl}$. Therefore, in order to have a model of inflation which is robust with respect to Planck corrections, the inflaton should be sub-Planckian, even if the additive Planck corrections can be suppressed.      

    However, should a large value of $r$ be observed, there is then the problem raised by the Lyth bound. The Lyth bound refers to the length of the inflaton trajectory. Therefore, if the inflaton potential can spiral in field space, for example in the space of a complex field $\Phi$,  then the bound can be satisfied while $|\Phi|$ remains sub-Planckian. Ideally, one would hope to find a model which is based on a straightforward particle physics effective theory and so is independent of assumptions about the nature of the complete theory of physics. 
There have been a number of effective field theory models with inflaton trajectories which can in some way spiral. (See, for example, \cite{knp,knp2,berg,bd1,ax0,ax1,nano,nano2,carone,baren}.)  
This was first achieved in \cite{knp} (see also \cite{knp2}), in a model based on two axions and two strongly-coupled gauge groups with aligned decay constants. In \cite{berg} a two-axion model was proposed with a single non-perturbative potential term plus an ad hoc quadratic potential for one of the axions. A similar scheme was proposed in \cite{bd1}, but with a second non-perturbative potential term for one of the axions replacing the quadratic potential term of \cite{berg}, which alleviates the fine-tuning required in the models of \cite{knp,knp2}. In \cite{carone} a variation of \cite{berg} with a symmetry-breaking potential was considered. SUGRA models of spiralling inflation were presented in \cite{nano,nano2}. 

   Recently, we proposed a minimal spiralling axion inflation model which is based on a single axion and a single strongly-coupled gauge group \cite{ax1}.  It effectively replaces the second axion of the two-axion models by $|\Phi|^{2}$, exploiting the natural coupling of $|\Phi|^2$ to $F \tilde{F}$. This model makes predictions which are quite different from those of the two-axion models, being dynamically equivalent to a $\phi^{4/3}$ chaotic inflation model in the absence of Planck corrections\footnote{An alternative model which also uses $|\Phi|$ and the phase of $\Phi$ was recently presented in \cite{baren}.}. 

    Here we will consider the effect of Planck corrections on the minimal sub-Planckian axion inflation model. In particular, in the case where the coupling of $|\Phi|^2$ to the topological charge density of the strongly-coupled gauge group has its dimensionally natural value, $\xi \sim 1$, $|\Phi|$ is close to $M_{Pl}$ and so Planck corrections can be significant. We will show that the effect of leading-order Planck  corrections is to significantly modify the tensor-to-scalar ratio without strongly altering the spectral index. Moreover, the shifts of the tensor-to-scalar ratio and the spectral index due to leading-order Planck corrections satisfy a consistency relation which is specific to the model. Observation of these shifts would therefore allow the model to be distinguished from a simple $\phi^{4/3}$ chaotic inflation model and at the same time provide a signature for the influence of Planck-scale physics on the inflaton potential.

\section{The Minimal Sub-Planckian Axion Inflation Model}

  The minimal sub-Planckian axion inflation model \cite{ax1} is structurally similar to the KSVZ axion model \cite{ksvz}. A complex field $\Phi$, the phase of which is the axion, couples to fermions $Q$ in the fundamental representation of a strongly-coupled gauge group,  
\be{e1} h \Phi \overline{Q}_{R} Q_{L} + h.c.    ~.\ee
The key difference from a conventional axion model is the direct coupling of $\Phi$ to the topological charge density of the strongly-coupled gauge group via the symmetry-invariant combination $\Phi^{\dagger}\Phi$,
\be{e2}  g^2 \xi \frac{|\Phi|^{2}}{M_{Pl}^{2}} F \tilde{F}  ~,\ee
where $\xi$ is a dimensionless parameter.  This coupling is consistent with all symmetries and would therefore be generally expected to exist as part of an effective theory. 
The quarks $Q$ are assumed gain a mass from the vacuum expectation value of $\Phi$, therefore the full renormalizable potential is  
\be{e3} V_{0}(\Phi) = -\mu^2 |\Phi|^2 + \lambda |\Phi|^4    ~.\ee
During inflation we will assume that the symmetry breaking mass term is negligible, therefore we will consider $V_{0}(\Phi) = \lambda |\Phi|^4$. As in the KSVZ model, $\Phi$ has charge $+1$ and $Q$ has charge $+1/2$ under a $U(1)_{A}$ global axial symmetry. In \eq{e1}, the phase $\theta$ of $\Phi \; (= \phi e^{i \theta}/\sqrt{2})$ can be rotated away via a local chiral transformation of the quarks. This results in a $U(1)_{A}$-breaking interaction of $\theta$ with the gauge fields due to the chiral anomaly,  
\be{e3a} \frac{g^2 \theta}{32 \pi^2} F \tilde{F}    ~.\ee
The total interaction with the gauge fields can therefore be written as 
\be{e5}  \frac{g^2}{32 \pi^2} \left( \frac{|\Phi|^{2}}{\Lambda^{2}} + \theta \right) F \tilde{F}   ~,\ee
where $\Lambda = M_{Pl}/(32 \pi^{2} \xi)^{1/2}$.
The quarks gain a mass from the large value of $|\Phi|$ during inflation. Under the assumption that the quarks are heavier than the strong-coupling scale $\Lambda_{sc}$, the resulting non-perturbative potential term generated by the strongly-coupled gauge sector is \cite{axrev} 
\be{e6}  V_{sc}(|\Phi|,\theta) = -\Lambda_{sc}^{4} \cos \left( \frac{|\Phi|^2}{\Lambda^2} + \theta \right)   ~.\ee
Therefore the full potential responsible for inflation is
\be{e7}  V(\Phi) = V_{0}(\Phi) + V_{sc}(\Phi) + \Lambda_{sc}^{4} = \lambda|\Phi|^4 + \Lambda_{sc}^{4} \left[1 - \cos \left( \frac{|\Phi|^2}{\Lambda^2} + \theta \right) \right]   ~,\ee
where we have added a constant term $\Lambda_{sc}^4$ so that the potential equals zero at the global minimum. 
For a range of values of $\Lambda_{sc}$ and $\xi$, this potential has a spiralling groove inscribed on the $|\Phi|^4$ potential in the complex plane\footnote{The handedness of the spiral is a result of the C- and CP-violating nature of the $F \tilde{F}$ term in \eq{e5}, which can produce a $U(1)_{A}$ charge asymmetry in the complex field.  
This can be seen since a field rotating in the complex plane has a global $U(1)$ charge density given by 
$\rho_{Q} = -i (\Phi^{\dagger} {\partial}_{0} \Phi - (\partial_{0} \Phi^{\dagger}) \Phi)= 2 \dot{\theta} |\Phi|^{2}$.}\cite{ax1}. Inflation occurs along this groove (very close to the angular or axion direction), allowing a super-Planckian change in the inflaton field while $|\Phi|$ remains sub-Planckian. The construction of the model is such that it makes quite specific predictions about the form of the potential, in particular the $|\Phi|^4$ perturbative potential and inclusion of $|\Phi|^2$ in the argument of the non-perturbative potential. (All other models so far proposed have only linear fields in the argument of the non-perturbative potential.) 

   It will be convenient to summarize the main results of \cite{ax1}. To a good approximation, the local minima of the potential along the $\phi$ direction satisfy 
\be{r1}  \frac{|\Phi|^{2}}{\Lambda^{2}}  \approx 2 n \pi - \theta ~,\ee
where $n$ is an integer. (This is demonstrated in Appendix A.) The direction of inflation is very close to the phase direction if $(d \phi/d \theta)^{2} \ll \phi^{2}$, and the resulting potential can be written in terms of an axion field $\hat{a}$ where $d \hat{a} = - \phi (d \theta/d \phi) d \phi$, $\phi = \sqrt{2} |\Phi|$ and $d \theta/d \phi$ is obtained from \eq{r1}. The effective inflaton $\ha$, which is defined such that $\ha = 0$ when $\phi = 0$, is related to $\phi$ by
\be{r2}  \ha = \frac{\phi^{3}}{3 \Lambda^{2}}    ~.\ee 
Therefore along the groove (where, as shown in Appendix A, the non-perturbative contribution to the potential is essentially equal to zero), $V \propto |\Phi|^4 \propto \hat{a}^{4/3}$ and so inflation along the axion direction is dynamically equivalent to a $\phi^{4/3}$ chaotic inflation model, which predicts $r = 16/3N = 0.097$ and $n_{s} = 1 - 5/3N = 0.970$ for $N = 55$.
$\ha$ and $|\Phi|$ are related to $N$ by 
\be{r3} \frac{\ha}{M_{Pl}} =  \left( \frac{8 N}{3} \right)^{1/2}     ~\ee 
and 
\be{r4} \frac{|\Phi|}{M_{Pl}} = \left( 3 N \right)^{1/6} 
\left( \frac{\Lambda}{M_{Pl}} \right)^{2/3}   ~.\ee
The value of $\Lambda$ is determined by the curvature perturbation at $N$ e-foldings
\be{r5}  \frac{\Lambda}{M_{Pl}} = \left( \frac{1}{3} \right)^{1/4} \left( \frac{1}{N} \right)^{5/8}  \left( \frac{8 \pi^2 P_{\zeta}}{\lambda} \right)^{3/8} ~.\ee
With $P_{\zeta}^{1/2} = 4.8 \times 10^{-5}$ and $N = 55$ this gives
\be{r6} \frac{\Lambda}{M_{Pl}} = 1.8 \times 10^{-4}\; \lambda^{-3/8}   ~\ee
and  so
\be{r7}  \frac{|\Phi|}{M_{Pl}} = 0.0075 \; \lambda^{-1/4}   ~.\ee
The corresponding value of $\xi$ is given by 
\be{r8} \xi = \frac{M_{Pl}^2}{32 \pi^2 \Lambda^2} = 9.8 \times 10^{4} \; \lambda^{3/4}     ~.\ee 
Therefore, in terms of $\xi$, 
\be{r9} \frac{|\Phi|}{M_{Pl}} = 0.35 \; \xi^{-1/3}    ~\ee
and 
\be{r10} \lambda = 2.2 \times 10^{-7} \; \xi^{4/3}    ~.\ee 
Thus $|\Phi| \lae 0.01M_{Pl}$ throughout $N = 60$ e-foldings when $\lambda \sim 1$ and $|\Phi| \lae 0.4 M_{Pl}$ when $\xi \sim 1$ \cite{ax1}.  

A number of conditions must be satisfied for the model to be consistent.  In order to have a local minimum in the radial direction close to the minimum of the non-perturbative potential, we require that $V_{sc}^{'}(\phi) \gg V^{'}_{0}(\phi)$ when $\phi$ significantly deviates from the minimum of $V_{sc}$.  This will be satisfied if $\Lambda_{sc} > \lambda^{1/4} (\Lambda \phi)^{1/2} = 1.4 \times 10^{-3} \lambda^{-1/16} M_{Pl}$. (Minimization in the radial direction is discussed further in Appendix A.) The effective mass squared in the radial direction at the local minimum is large enough to reduce the dynamics of the model to a single-field model in the $\ha$ direction if $V_{sc}^{''}(\phi) \gg H^2$, which is satisfied if $\Lambda_{sc} > 1.0 \times 10^{-5} \; \lambda^{-1/4}  M_{Pl}$. The axion will be very close to the angular direction if $ (d \phi/d \theta)^{2} \ll \phi^{2}$. This is satisfied if $(\Lambda/\phi)^{4} \ll 1$, which requires that $\lambda > 7 \times 10^{-15}$ and is therefore easily satisfied. 
Finally, the form of the non-perturbative potential is correct if the quark masses are large compared to $\Lambda_{sc}$, otherwise $\Lambda_{sc}^{4} \rightarrow m_{Q} \Lambda_{sc}^{3}$ \cite{axrev}. With $m_{Q} = h |\Phi| \sim |\Phi|$, this requires that\footnote{If this condition were violated it would not prevent inflation, it would only alter the form of the non-perturbative potential. In general, increasing $\Lambda_{sc}$ strengthens the assumptions leading to inflation.}    $\Lambda_{sc} \lae |\Phi|$ and so $\Lambda_{sc} \lae 7.5 \times 10^{-3} \lambda^{-1/4} M_{Pl}$. In particular, for $\xi = 1$ and $\lambda = 2.2 \times 10^{-7}$, all conditions are satisfied if $0.004 M_{Pl} \lae  \Lambda_{sc} \lae 0.4 M_{Pl}$.

    In \cite{ax1} we focused on two cases of interest. In one case we considered $\lambda$ to have its dimensionally natural value, $\lambda \sim 1$, in which case $|\Phi|/M_{Pl} \lae 0.01$ during inflation. In this case the additive Planck corrections to the potential, $g(|\Phi|)$, are small enough relative to $V(\Phi)$ to have no significant effect on inflation. Therefore this model completely solves the Planck correction problem, but it does so at the cost of a large dimensionless coupling, $\xi \sim 10^5$. The other case we considered is where $\xi \sim 1$. In this case, inflation requires that $\lambda \sim 10^{-7}$. This is a much less severe bound than in the case of conventional chaotic inflation models, where the $\phi$ self-coupling must satisfy $\lambda \lae 10^{-14}$. The field remains sub-Planckian, but it is now much closer to the Planck scale, with $|\Phi|/M_{Pl} = 0.35$ for $\xi = 1$. Therefore the model does not automatically solve the problem of additive Planck corrections to the potential, since a $|\Phi|^{6}/M_{Pl}^4$ term would dominate the $\lambda |\Phi|^4$ potential\footnote{The general condition for the contribution of the $|\Phi|^{6}/M_{Pl}^{4}$ term to the $\phi$ field equation to be smaller than that of the $\lambda |\Phi|^4$ term is $\lambda > 2 \times 10^{-3}$ and $\xi > 900 $.}.  One must therefore assume that the additive Planck corrections are also suppressed by at least a factor of $\lambda$. However, this is a much smaller suppression than would be required if the field were super-Planckian during inflation. This suppression is not implausible if the renormalizable and non-renormalizable potential terms have a common origin in the complete theory, or if $\lambda$ represents a symmetry-breaking effect such that the complete potential is exactly zero in the limit $\lambda \rightarrow 0$. However, even if the additive corrections are suppressed, in the absence of a complete theory  we should generally expect multiplicative Planck corrections. The sub-Planckian value of $|\Phi|$ predicted by the model allows such corrections to be controlled and their effects to be understood via an expansion in $|\Phi|^{2}/M_{Pl}^{2}$. 

\section{Modification of Predictions due to Planck Corrections} 

   In general, we should expect additive and multiplicative Planck corrections of the form\footnote{The fields are defined throughout to be canonically normalized, in which case Planck corrections appear only in the potential and interaction terms. We have not included derivatives of the field $\phi$ in the Planck corrections. Derivative corrections will be small compared to those involving only the field itself if 
$\dot{\phi}/\phi \sim H \ll \phi$, which is generally true in the model we are considering.},  
\be{v1}  V_{0}(|\Phi|) = \lambda |\Phi|^4 \rightarrow  \lambda f_{v}\left(
\frac{|\Phi|^{2}}{M_{Pl}^2}\right) |\Phi|^{4} +  g_{v}\left(\frac{|\Phi|^{2}}{M_{Pl}^2}\right)|\Phi|^{4}  ~\ee
and 
\be{v2} g^{2} \xi \frac{|\Phi|^{2}}{M_{Pl}^{2}}F \tilde{F} \rightarrow   g^2 \left[f_{\xi}\left(
\frac{|\Phi|^{2}}{M_{Pl}^{2}}\right) 
\frac{\xi |\Phi|^{2}}{M_{Pl}^{2}}  
+ g_{\xi}\left(\frac{|\Phi|^{2}}{M_{Pl}^{2}}\right)  \right]F \tilde{F} ~.\ee
If $|\Phi|^{2}/M_{Pl}^{2}$ is small compared to 1, we can expand $f_{i}$ and $g_{i}$ as 
\be{v3}  f_{i} = \sum_{r = 0}^{\infty} \alpha_{i_{r}} \left(\frac{|\Phi|^{2}}{M_{Pl}^{2}}\right)^{r}  \;\;\; ; \;\;\;\;  g_{i} = \sum_{r = 1}^{\infty}  \beta_{i_{r}}\left(\frac{|\Phi|^{2}}{M_{Pl}^{2}}\right)^{r}  ~,\ee
where $i = v$ or $\xi$ and we define $\alpha_{i_{0}} = 1$.

In the case where $g_{v}$ is suppressed by a factor of $\lambda$ per term\footnote{Note that this would provide a sufficient suppression of the Planck corrections for sub-Planckian values of $|\Phi|$ but not for super-Planckian values.}, the additive correction becomes effectively a multiplicative correction to the $\lambda |\Phi|^4$ potential, therefore we can set $g_{v} = 0$. In addition, if $\xi \sim 1$, the $|\Phi|$-dependent terms from the additive correction to the $F \tilde{F}$ coupling are of the same magnitude as those from the first term and so we can set $g_{\xi} = 0$.  To simplify the notation, we expand $f_{v}$ and $f_{\xi}$ as 
\be{v5} f_{v} = 1 + a_{1} \frac{|\Phi|^{2}}{M_{Pl}^{2}} + a_{2} \frac{|\Phi|^{4}}{M_{Pl}^{4}} + ...   ~\ee
and 
\be{v6} f_{\xi} = 1 + b_{1} \frac{|\Phi|^{2}}{M_{Pl}^{2}} + b_{2} \frac{|\Phi|^{4}}{M_{Pl}^{4}} + ...   ~.\ee
In practice, we will consider only the leading-order corrections, since the higher-order corrections only become important once the sub-Planckian expansion is already breaking down. Dimensionally we expect $|a_{1}| \sim |b_{1}| \sim 1$.

    We next consider how the leading-order Planck corrections modify the predictions of the model. We first consider the 
$a_{1}$ correction. In this case the inflaton potential becomes 
\be{v7}  V_{0}(\Phi) = \lambda \left(|\Phi|^{4} + a_{1} \frac{|\Phi|^{6}}{M_{Pl}^{2}} \right)   ~.\ee
As in the original model, at the minimum of the non-perturbative potential $|\Phi|$ is related to the phase $\theta$ by \eq{r1}. The direction of inflation is very close to the phase direction, and the resulting potential can be written in terms of the axion field $\hat{a}$, which is related to $\phi$ by \eq{r2}.
Substituting $\phi = (3 \Lambda^{2} \ha)^{1/3}$ into the potential \eq{v7}, we obtain 
\be{v10} V(\hat{a}) = \frac{\lambda}{4} K^{4} \hat{a}^{4/3} + \frac{\lambda a_{1}}{8} \frac{K^{6} \hat{a}^{2}}{M_{Pl}^{2}}   ~,\ee
where we have defined $K = (3 \Lambda^{2})^{1/3} $. The form of the leading-order correction is quite specific to the minimal sub-Planckian axion inflation model.

The leading-order Planck correction to the potential modifies $N$, $\eta$ and $\epsilon$  as a function of $\ha$. To leading order in an expansion in $\ha^{2}/M_{Pl}^{2}$,
\be{v11} N = \frac{1}{M_{Pl}^{2}} \left[ \frac{3}{8} \ha^2 - \frac{9 a_{1}}{128} \frac{K^{2}}{M_{Pl}^{2}} \ha^{8/3} \right] ~,\ee
\be{v12} \eta = \frac{4}{9} \frac{M_{Pl}^{2}}{\ha^{2}} \left(1 + \frac{7 a_{1}}{4}  \frac{K^{2}}{M_{Pl}^{2}} \ha^{2/3} \right)   ~\ee 
and 
\be{v13} \epsilon =  \frac{8}{9} \frac{M_{Pl}^{2}}{\ha^{2}} \left(1 + \frac{a_{1}}{2}\frac{K^{2}}{M_{Pl}^{2}} \ha^{2/3} \right)   ~.\ee      
Inverting \eq{v11} gives $\ha^{2}$ as a function of $N$, 
\be{v14} \frac{\ha^{2}}{M_{Pl}^{2}} = \frac{8}{3} N \left( 1 + \frac{3 a_{1}}{16} \left(\frac{8}{3}\right)^{1/3} \frac{K^{2}N^{1/3} }{M_{Pl}^{4/3}} 
\right)  ~.\ee 
It is useful to express the leading-order corrections to $\epsilon$ and $\eta$ in terms of $|\Phi|^{2}/M_{Pl}^{2}$ at $N$ e-foldings. We explicitly show this for the case of $\eta$ in Appendix B. We find   
\be{v15} \eta = \frac{1}{6N} \left(1 + \frac{25 a_{1}}{8} \frac{|\Phi|^{2}}{M_{Pl}^{2}} \right)   ~\ee 
and 
\be{v16} \epsilon = \frac{1}{3N} \left(1 + \frac{5 a_{1}}{8} \frac{|\Phi|^{2}}{M_{Pl}^{2}} \right)   ~.\ee 
Using the standard relations $n_{s} = 1 + 2 \eta - 6 \epsilon$ and $r = 16 \epsilon$ we then obtain 
\be{v17} n_{s} = 1 - \frac{5}{3 N} \left(1 + \frac{a_{1}}{8} \frac{|\Phi|^{2}}{M_{Pl}^{2}} \right)  ~\ee 
and 
\be{v21}  r = \frac{16}{3N} \left( 1 + \frac{5 a_{1}}{8} 
\frac{|\Phi|^{2}}{M_{Pl}^{2}}\right) ~.\ee
These imply a consistency relation between the shift of $n_{s}$ and the shift of $r$ relative to the predictions of the uncorrected model, which are equivalent to those of a $\phi^{4/3}$ chaotic inflation model. Defining the unshifted values to be $r_{0} = 16/3N$ and $n_{s\;0} = 1 - 5/3N$, we find
\be{v22} \Delta r \equiv r - r_{0} = \frac{10 a_{1}}{3 N} 
\frac{|\Phi|^{2}}{M_{Pl}^{2}}   ~\ee
and 
\be{v23}  \Delta n_{s} \equiv n_{s} - n_{s\;0} = - \frac{5 a_{1}}{24 N} 
\frac{|\Phi|^{2}}{M_{Pl}^{2}}    ~.\ee
Therefore 
\be{v23a} \Delta n_{s} = - \frac{\Delta r}{16}    ~.\ee
This relation is specific to the potential \eq{v10} and is therefore specific to the minimal sub-Planckian axion inflation model. 

We next consider the $b_{1}$ correction. In this case there are no Planck corrections to the potential but the relation between $|\Phi|$ and $\ha$ is altered. The non-perturbative potential is modified to 
\be{v24}  V_{sc} = \Lambda_{sc}^{4} \left[1 - \cos \left( \frac{|\Phi|^{2}}{\Lambda^{2}} + \frac{b_{1}}{\Lambda^{2}} \frac{|\Phi|^{4}}{M_{Pl}^{2}} + \theta \right) \right]   ~.\ee 
Therefore at the minimum 
\be{v25} \theta = 2 n \pi -  
\frac{|\Phi|^{2}}{\Lambda^{2}} - \frac{b_{1}}{\Lambda^{2}} \frac{|\Phi|^{4}}{M_{Pl}^{2}}  ~.\ee
Thus 
\be{v27} \frac{d \theta}{d \phi} = - \left( 
\frac{\phi}{\Lambda^{2}} + \frac{b_{1} \phi^{3}}{\Lambda^{2} M_{Pl}^{2}} \right)  ~.\ee
Therefore  
\be{v26} \ha = \int d \ha = -\int \phi \left(\frac{d \theta}{d \phi} \right) d \phi = \frac{\phi^{3}}{3 \Lambda^{2}}  \left(1 + \frac{3 b_{1}}{5} \frac{\phi^{2}}{M_{Pl}^{2}} \right)   ~.\ee 
Inverting this to first-order gives 
\be{v29} \phi = \left(3 \Lambda^{2}\right)^{1/3} \left( \ha^{1/3} - \frac{3^{2/3} b_{1}}{5} \frac{\Lambda^{4/3} \ha}{M_{Pl}^{2}}  \right)   ~.\ee 
Substituting this into the potential $V(\phi) = \lambda \phi^{4}/4$ then gives the effective potential for $\ha$, 
\be{v30} V = \frac{\lambda}{4} K^{4} \ha^{4/3} - \frac{\lambda b_{1}}{5} \frac{K^{6}  \ha^{2}}{M_{Pl}^{2}}     ~.\ee 
By comparing with \eq{v10}, the $b_{1}$ correction to the potential can be seen to be equivalent to the $a_{1}$ correction with $a_{1} = -8 b_{1}/5$. Therefore, to leading order in Planck corrections, we can use the results for $a_{1}$ with $a_{1} \rightarrow \tilde{a}_{1} = a_{1} - 8 b_{1}/5$. Thus in the case where $b_{1} \sim -a_{1}$, the combined effect of the $a_{1}$ and $b_{1}$ corrections would be to produce a significantly larger effect on $n_{s}$ and $r$ at a given value of $|\Phi|^{2}/M_{Pl}^{2}$ than would otherwise be the case.

\section{Results}

    We are interested in the effect of Planck corrections when the expansion in terms of $|\Phi|^{2}/M_{Pl}^{2}$ is valid. To be definite, we will define this to be valid when $|\Phi|^{2}/M_{Pl}^{2} \leq 0.25$, assuming that $|a_{1}|, \;|b_{1}| \leq 1$. We will refer to this as the sub-Planckian regime. The shifts in $n_{s}$ and $r$ for $N = 55$ are then given by
\be{v31} \Delta n_{s} = -9.5 \times 10^{-4} \ta_{1} \left(\frac{|\Phi|^{2}/M_{Pl}^{2}}{0.25}\right) \;\;\;;\;\;\; \Delta r = 0.015 \; \ta_{1} \left(\frac{|\Phi|^{2}/M_{Pl}^{2}}{0.25}\right)    ~\ee
In the case where the $a_{1}$ correction is dominant and $|a_{1}| \lae 1$, we find $|\Delta r| \lae 0.02$ and $|\Delta n_{s}| \lae 0.001$.  Thus it is possible to obtain a significant shift of $r$ within the sub-Planckian regime, whereas the shift of $n_{s}$ is generally small.
The shifts of $n_{s}$ and $r$ due to leading-order Planck corrections are correlated by \eq{v23a}, allowing the model to be tested if specific values of $n_{s}$ and $r$ can be obtained with sufficiently accuracy. 
For $|a_{1}| \lae 1$  the model predicts $(n_{s}, r)$ to be in the range $(0.969, 0.12)$ to $(0.971, 0.08)$ when $N = 55$. (The most recent Planck bounds are $r < 0.11$ (2-$\sigma$) and $n_{s} = 0.9677 \pm 0.0060$ (1-$\sigma$) (Planck TT + lowP + lensing) \cite{planckcos}). The value of $n_{s}$ therefore  largely preserves the $\phi^{4/3}$ chaotic inflation prediction, $n_{s} = 0.970$. 
 
In the case where both corrections are present, with $a_{1} \sim - b_{1}$ and $|a_{1}| \lae 1$, we can have a larger $\ta_{1}$ for a given value of $|\Phi|/M_{Pl}$ within the sub-Planckian regime, 
with $|\ta_{1}| \sim 3$ being possible. In this case $|\Delta n_{s}| \sim 0.003$ and $|\Delta r| \sim 0.06$ is possible. The predicted range of  $(n_{s}, r)$ is then $(0.967, 0.16)$ to $(0.973, 0.04)$, with the shifts again correlated by \eq{v23a}.

\section{Conclusions} 

  We have considered the effect of leading-order Planck corrections on the minimal sub-Planckian axion inflation model. For values of $|\Phi|/M_{Pl}$ which are small enough to allow the corrections to be understood in an expansion in $|\Phi|^{2}/M_{Pl}^{2}$, Planck corrections can substantially shift the value of $r$ from its uncorrected value of $r = 0.097$ (using $N = 55$ for the Planck pivot), with 0.04 to 0.16 being possible. The possibility of reducing $r$ from its uncorrected value may be significant, given that the current 2-$\sigma$ upper bound from Planck is $r < 0.11$ \cite{planckcos}. The corresponding shift of $n_{s}$ is much smaller, with values in the range 0.967 to 0.973. This substantially retains the $n_{s}$ prediction of the uncorrected model (equivalent to a $\phi^{4/3}$ chaotic inflation model), $n_{s} = 0.970$. The corresponding Planck result is $n_{s} = 0.9677 \pm 0.0060$, which can accommodate the expected range of $n_{s}$ within 1-$\sigma$.

     In addition, there is a consistency relation between the shifts of $n_{s}$ and $r$ due to leading-order Planck corrections, which is specific to the minimal spiralling axion inflation model, $\Delta n_{s} = -\Delta r/16$. This simultaneously provides a test of the model, a means to distinguish the model from a $\phi^{4/3}$ chaotic inflation model and a signature for the influence of Planck corrections.

   The value of $N$ corresponding to the Planck pivot scale depends on the reheating temperature and is not precisely known at present. In the absence of a precise value for $N$, the observed values of $n_{s}$ and $r$ can be used to fix the value of $N$ and the coefficient of the leading-order Planck correction. These can then be checked for consistency with a physically reasonable value for $N$ and a sub-Planckian value for the magnitude of the correction. In contrast, the predictions of the $\phi^{4/3}$ chaotic inflation model depend only on $N$ and so the $\phi^{4/3}$ chaotic inflation model can be excluded by values of $n_{s}$ and $r$ which are nevertheless consistent with the Planck-corrected spiralling axion inflation model. It may also be possible to determine the reheating temperature and so $N$ in the spiralling axion inflation model, given that it is based on an effective particle physics theory. We will return to this issue in future work.

\section*{Acknowledgements} This work was partially supported  by STFC grant ST/J000418/1.

\renewcommand{\theequation}{A-\arabic{equation}}
 \setcounter{equation}{0} 

\section*{Appendix A: Minimization of the potential in the radial direction}

   Here we show that the condition $\Lambda_{sc} > \lambda^{1/4}(\phi_{0}\Lambda)^{1/2}$ ensures that the minimum of the potential in the radial direction is close to the minimum \eq{r1} of the non-perturbative potential and that the shift of the potential $\Delta V$ due to the shift $\Delta \phi$ from the minimum of the non-perturbative potential satisfies $|\Delta V/V_{0}| \ll 1$. 

   The condition that the shift of $\phi$ from the minimum \eq{r1} is small enough to be negligible is that the shift of the argument of the non-perturbative potential is small compared to the spacing $2 \pi$ of the minima. We will therefore require that $\Delta~|\Phi|^{2}/\Lambda^{2}~\ll~1$. To compute the shift $\Delta \phi$, we expand the potential \eq{e7} about a minimum $\phi_{0}$ of the non-perturbative potential. This gives 
\be{bb1} \Delta V = \lambda \phi_{0}^{3} \Delta \phi + \frac{1}{2} \left(\frac{\Lambda_{sc}^{4}}{\Lambda^{4}} + 3 \lambda\right)  \phi_{0}^{2} \Delta \phi^{2}    ~.\ee
In general we find that $\Lambda_{sc} > \Lambda$, therefore we can neglect the $3 \lambda$ term. Minimizing \eq{bb1} then gives 
\be{bb2} \Delta \phi = - \frac{\lambda \phi_{0} \Lambda^{4}}{\Lambda_{sc}^{4}}    ~.\ee
The shift of the argument of the non-perturbative potential is given by 
\be{bb3} \Delta \left(\frac{|\Phi|^{2}}{\Lambda^{2}}\right) = \frac{\phi_{0} \Delta \phi}{\Lambda^{2}}  ~.\ee
Requiring that this is much smaller than 1 then requires that 
$|\Delta \phi| \ll \Lambda^{2}/\phi_{0}$. From \eq{bb2}, this is satisfied if $\Lambda_{sc} < \lambda^{1/4}(\phi_{0} \Lambda)^{1/2}$. The same condition can be obtained by requiring that $V^{'}_{sc}(\phi) \gg V^{'}_{0}(\phi)$ when the argument of the non-perturbative potential is such that $\sin (|\Phi|^{2}/\Lambda^{2} + \theta) \sim 1$ \cite{ax1}.  

     We next check that the shift of the potential $\Delta V$ due to $\Delta \phi$ is negligible. Substituting \eq{bb2} into \eq{bb1}, we obtain
\be{bb4} \Delta V = - \frac{\lambda^{2} \phi_{0}^{4}}{2} \left( 
\frac{\Lambda}{\Lambda_{sc}}\right)^{4}   ~.\ee
Therefore, with $V_{0} = \lambda \phi_{0}^{4}/4$ and $\Lambda_{sc} > \lambda^{1/4} (\Lambda \phi_{0})^{1/2}$, we find that
\be{bb5} \frac{|\Delta V|}{V_{0}} \ll  \frac{2 \Lambda^{2}}{\phi_{0}^{2}} =  \frac{\Lambda^{2}}{|\Phi_{0}|^{2}}  ~.\ee
Using \eq{r6} and \eq{r7} we then obtain 
\be{bb6} \frac{|\Delta V|}{V_{0}} \ll 0.0006 \; \lambda^{-1/4}   ~.\ee
With $\lambda \sim 10^{-7}$ when $\xi \sim 1$, this implies that $|\Delta V|/V_{0} \ll 0.03$. 

Therefore it is a good approximation to assume that $\phi$ is exactly at the minimum of the non-perturbative potential and that $V$ at the minimum is equal to $V_{0} = \lambda \phi^{4}/4$.

\renewcommand{\theequation}{B-\arabic{equation}}
 \setcounter{equation}{0} 

\section*{Appendix B: Leading-order Planck correction to $\eta$}

Here we explain the calculation of the leading-order correction to $\eta$ in terms of $|\Phi|^{2}/M_{Pl}^{2}$. (The calculation of $\epsilon$ is similar.) In terms of $\ha$, $\eta$ is given by 
\be{aa1} \eta = \frac{4}{9} \frac{M_{Pl}^{2}}{\ha^{2}} \left(1 + \frac{7 a_{1}}{4}  \frac{K^{2}}{M_{Pl}^{2}} \ha^{2/3} \right)   ~.\ee 
From \eq{v14}, 
\be{aa1a} \frac{\ha^{2}}{M_{Pl}^{2}} = \frac{8}{3} N \left( 1 + \frac{3 a_{1}}{16} \left(\frac{8}{3}\right)^{1/3} \frac{K^{2}N^{1/3} }{M_{Pl}^{4/3}} 
\right)   =   \frac{8}{3} \left(N + \frac{9 a_{1}}{128} \frac{K^{2}}{M_{Pl}^{4}} \ha^{8/3} \right)  ~,\ee
where we have used the zeroth-order relation $\ha^{2}/M_{Pl}^{2} = 8N/3$ in the second term.  Substituting this into the prefactor of \eq{aa1} gives
\be{aa2} \eta = \frac{4}{9} \times
\frac{1}{
\frac{8}{3} \left(N + \frac{9 a_{1}}{128} \frac{K^{2}}{M_{Pl}^{4}} \ha^{8/3} \right)  } \times  \left(1 + \frac{7 a_{1}}{4}  \frac{K^{2}}{M_{Pl}^{2}} \ha^{2/3} \right)  ~.\ee
We then expand the denominator to first-order in the Planck correction, which gives
\be{aa3} \eta = \frac{1}{6N} \left( 1 - \frac{9}{128} \frac{a_{1} K^{2}}{M_{Pl}^{4}N} \ha^{8/3} +   \frac{7 a_{1}}{4}  \frac{K^{2}}{M_{Pl}^{2}} \ha^{2/3} \right)   ~.\ee 
Since the last two terms are first-order Planck corrections, we can substitute the relation between $\ha$ and $|\Phi|$, $\ha^{2/3} = 2 |\Phi|^{2}/K^{2}$, which follows from \eq{r2}, and the zeroth-order relation between $|\Phi|/M_{Pl}$ and $N$, $|\Phi|/M_{Pl} = (3 N)^{1/6}(\Lambda/M_{Pl})^{2/3}$, to obtain the final expression 
\be{aa4} \eta = \frac{1}{6N} \left(1 + \frac{25}{8} a_{1} \frac{|\Phi|^{2}}{M_{Pl}^{2}} \right)   ~.\ee

\end{document}